\documentclass[preprint,pre,aps]{revtex4}
\usepackage{graphics}
\usepackage{amsmath,amssymb}
\usepackage[dvips]{graphicx}
\begin{document}
\title{The Ginzburg temperature of ionic fluids revisited}
 \author{O. Patsahan}
\affiliation{Institute for Condensed Matter Physics of the National
Academy of Sciences of Ukraine, 1 Svientsitskii Str., 79011 Lviv,
Ukraine}

\date{\today}
\begin{abstract}
Using the collective variables method  we revisit the estimates of
the Ginzburg temperature  for the Coulomb-dominated models of ionic
fluids. We consider the  charge-asymmetric primitive model
supplemented by  short-range attractive interactions in the vicinity
of the gas-liquid critical point. For this model, we derive the
effective Ising-like $\phi^{4}$-model Hamiltonian expressed in terms
of the collective variables describing the fluctuation modes of the
total number density.  We obtain the explicit expressions for all
the Hamiltonian coefficients within the framework of the same
approximation, namely, the one-loop approximation which produces the
mean-field critical parameters. Based on this Hamiltonian, we
consistently calculate the reduced Ginzburg temperature $t_{G}$ for
both the pure Coulombic model (a restricted primitive model) and the
pure solvophobic model (a hard-sphere square-well model) as well as
for the model parameters ranging between these two limiting cases.
Contrary to the previous theoretical estimates, we obtain the
reduced Ginzburg temperature for the pure Coulombic model to be
about $20$ times smaller than for the pure solvophobic model. For
the full model including the short-range attractive and  long-range
Coulomb interactions, we show that $t_{G}$   approaches
the value found for the pure Coulombic model when the strength of
the Coulomb interactions becomes large enough. Our results suggest a
key role of the Coulomb interactions in the crossover  behaviour
observed experimentally in ionic fluids as well as  confirm the
Ising-like criticality in the Coulomb-dominated ionic systems.
\end{abstract}

\maketitle
\section{Introduction}
For the last two decades, much attention has been focused on the
issue of  critical and phase behaviour of fluids with Coulomb
interactions. These studies were motivated by  controversial
experimental results that demonstrated the three types of the
critical behaviour in ionic fluids:  Ising-like critical behaviour,
classical mean-field behaviour, and  a crossover between these two
regimes (see Refs~\cite{Buback-Franck,Chieux-Sienko,singh,levelt1}).
In order to interpret the results, the ionic systems were classified
as either solvophobic or Coulombic. The main attention has been paid
to the criticality of  Coulombic systems in which the phase
separation is primarily driven by long-range electrostatic
interactions.

It is now generally accepted that  the critical behaviour  of
Coulombic systems belongs to the universality class of a
three-dimensional Ising model. Earlier experiments that supported
the expectation of mean-field critical behaviour could not be
reproduced in later works (see Ref.~\cite{Schroer:review,Schroer:12}
and references therein). Precise experiments
indicate  a crossover from Ising to mean-field behaviour
characterized by an increase of the non-classical region with the
polarity increase of the solvent \cite{Schroer:12,
Sengers_Shanks:09}. 

Many theoretical and numerical works on  Coulombic systems are based
on a restricted primitive model (RPM), i.e., an equimolar mixture of
equisized  charged hard spheres immersed in a structureless
dielectric continuum. It is established that the RPM undergoes a
gas-liquid-like phase transition at low temperature and low density
\cite{stell1,levin-fisher,Cai-Mol1,patsahan_ion}. Reliable estimates
of  the  location of the critical point have been obtained using
simulations \cite{caillol-levesque-weis:02,Hynnien-Panagiotopoulos}.
Simulations also strongly support the Ising critical behaviour of
the model
\cite{caillol-levesque-weis:02,Hynnien-Panagiotopoulos,luijten,kim-fisher-panagiotopoulos:05}.

The major part of  theoretical studies of the criticality in the RPM
is based on the mean-field  theories and deals with calculations of
the reduced Ginzburg temperature $t_{G}$ \cite{levanyuk,ginzburg}.
According to the Ginzburg criterion \cite{Goldenfeld},  the
mean-field theory is valid only when $|t|\gg t_{G}$, where
$t=(T-T_{c})/T_{c}$ and $T_{c}$ are the mean-field reduced
temperature and the mean-field critical temperature, respectively. 
Generally, the previous estimates of the reduced Ginzburg
temperature of the RPM suggest the non-classical region in the
Coulombic systems to be of the same order, or even larger than in
simple fluids \cite{evans,fisher3,schroer}. Thus,  excluding the
possibility of mean-field behavior of the RPM, these results fail to
explain the experimentally observed  reduction of the crossover
temperature in  Coulombic systems.  However, a very small value of
$t_{G}$ for the RPM (about  $10^{3}$ times smaller than for a simple
fluid model)  was found in Ref.~\cite{schroer_1}. It should be stressed
that two conditions are to be taken into account  to get a
reasonable result for the Ginzburg temperature. The fist condition
consists in the choice of ``a reference model`` exhibiting typical
Ising critical behaviour. The second one is the necessity to
consider both the reference model and the  model studied  at the
same level of approximation. In Refs.~\cite{evans,fisher3,schroer},
the values of the Ginzburg temperature  were calibrated by a typical
model of simple fluids. However, they were  mainly calculated at
different, although often comparable, levels of approximation.

On the other hand, the criticality of  Coulombic systems has been
studied using the functional integration methods
\cite{ciach:06:1,brilliantov,patsahan:04:1,patsahan-caillol-mryglod:07,moreira-degama-fisher}.
In particular, several attempts have been made in order to derive
the effective Ginzburg-Landau-Wilson (GLW) Hamiltonian of the RPM
\cite{brilliantov,patsahan:04:1}. In Ref.~\cite{brilliantov}, the
effective Hamiltonian is obtained in terms of a scalar field
conjugate to the charge density. However, a non-perturbative
renormalization group analysis of such Hamiltonian does not allow
one to make an unambiguous statement on the nature of the critical
behaviour of the RPM.   An attempt to derive the effective
Hamiltonian of the RPM in terms of  the number density field, strong
fluctuating quantity in the vicinity of the gas-liquid critical
point, was made using the collective variables (CVs) method in
Ref.~\cite{patsahan:04:1}. The analysis of the Hamiltonian
coefficients shows that in spite of the long-range character of the
Coulomb potential, the effective interactions are of a short-range
character and describe attraction. The form of the Hamiltonian
suggests the Ising-like criticality  of the  RPM. However, the
numerical estimations of the relevant coefficients   were not
presented in this work. More recently, an effect of  the long-range
interactions on the Ginzburg temperature has been studied on the
basis of the LGW Hamiltonian expressed in terms of the  field
conjugate to the order parameter \cite{patsahan-caillol-mryglod:07}.
The Hamiltonian coefficients are presented therein in the form of an
expansion in powers of the ionicity measuring the strength of the
Coulomb interaction, and the consideration is restricted to the
second power. The results have shown that an increase in the Coulomb
interactions leads to a decrease of the temperature region of the
crossover regime which confirms the experimental observations
\cite{Kleemeier}. A  reduction of the reduced crossover temperature
with an increase of the ionicity is also indicated in
Ref.~\cite{moreira-degama-fisher}, though it is weak  compared to
the experimental data. The limiting system with the pure Coulomb
interactions, which is the RPM,  lies outside the range of validity
of the perturbative treatments developed in
Refs~\cite{patsahan-caillol-mryglod:07,moreira-degama-fisher}.

Recently, non-classical critical exponents have been found for the RPM using 
the hierarchical reference theory \cite{Parola-Reatto:11}. However, 
an issue of the width of the  critical region has not been addressed in this work.    

Summarizing, we can state that  the nature of the non-classical region in the Coulomb
dominated systems remains of fundamental interest and presents a
real challenge.

The purpose of the present paper is to derive a microscopic-based
effective Hamiltonian of the  ionic model supplemented by
short-range  attractive interactions and on its basis to
consistently calculate   the Ginzburg temperature for  both the pure
Coulombic model  and  the pure solvophobic model  as well as for the
model parameters  ranging between the two limiting cases. To this
end, we use the CVs-based theory developed for the description of
phase transitions in ionic systems (see
Refs.~\cite{Pat-Mryg-CM,patsahan-mryglod-patsahan:06}). Following
the ideas of Ref.~\cite{patsahan:04:1}, we integrate out  the
variables connected with the charge-density fluctuations and derive
the effective Hamiltonian in terms of the variables  describing the
total number density fluctuations. In this paper, we find the
explicit expressions for all the  coefficients of the effective
$\phi^{4}$-model  Hamiltonian at the same level of  approximation,
namely, in the one-loop approximation corresponding to a one sum
over the wave vector. This enables us to get  consistent estimates
for  the critical parameters as well as for the Hamiltonian
coefficients for a whole range of the model parameters. The Ginzburg
temperature for the pure Coulombic model obtained in this way
appears to be  about twenty times smaller than for a simple fluid.
We also study the effect of the interplay of short-range and
long-range interactions on the Ginzburg temperature.

The paper is arranged as follows. In Sec.~II we give some brief
background to the CVs-based theory for a charge-asymmetric primitive
model  with additional short-range attractive interactions included.
Sec.~III is devoted to the derivation  of the effective Hamiltonian
of the model in the vicinity of the gas-liquid critical point. In
Sec.~IV we calculate the Ginzburg temperature for the hard-sphere
square-well model, the RPM  as well as for the  models including
both short-range and long-range interactions. We conclude in Sec.~V.

\section{Background}
We start with  a classical two-component system
consisting of $N_{1}$ particles of species $1$ and $N_{2}$ particles
of species $2$. The pair interaction potential is assumed to be of
the following form:
\begin{equation}
U_{\alpha\beta}(r)=\phi_{\alpha\beta}^{HS}(r)+\phi_{\alpha\beta}^{C}(r)+\phi_{\alpha\beta}^{SR}(r),
\label{2_1}
\end{equation}
where $\phi_{\alpha\beta}^{HS}(r)$ is the interaction potential
between the two additive hard spheres of diameters $\sigma_{\alpha}$
and $\sigma_{\beta}$.  Here, $\phi_{\alpha\beta}^{C}(r)$ is the
Coulomb potential:
$\phi_{\alpha\beta}^{C}(r)=q_{\alpha}q_{\beta}\phi^{C}(r)$, where
$\phi^{C}(r)=1/(\epsilon r)$, $\epsilon$ is the dielectric constant.
The system consists of both positive and negative ions so that the
electroneutrality condition is satisfied, i.e.,
$\sum_{\alpha=1}^{2}q_{\alpha}\rho_{\alpha}=0$, where
$\rho_{\alpha}$ is the number density of species $\alpha$,
$\rho_{\alpha}=N_{\alpha}/V$, $V$ is the volume of the system. The
ions of the species $\alpha=1$ are characterized by their
hard-sphere diameter $\sigma_{1}$ and by their electrostatic charge
$q_{0}$ and those of species $\alpha=2$ are characterized by
diameter $\sigma_{2}$ and  opposite charge $-zq_{0}$ ($q_{0}$ is an
elementary charge and $z$ is the parameter of charge asymmetry).
Hereafter we consider the case  $\sigma_{1}=\sigma_{2}=\sigma$. The
potential $\phi_{\alpha\beta}^{SR}(r)$ describes the short-range
attraction. We specify   $\phi_{\alpha\beta}^{SR}(r)$ in the form of
the square-well (SW) potential of the range $\lambda$ and assume
$\phi_{11}^{SR}(r)=\phi_{22}^{SR}(r)=\phi_{12}^{SR}(r)=\phi^{SR}(r)$.
The system of hard spheres interacting through the SW potential with
$\lambda=1.5\sigma$ can serve as a reasonable model for simple
fluids. Such a system   undergoes a gas-liquid critical point which belongs to
the universal class of a three-dimensional Ising model.

Using the CVs method, we can present the functional of the grand partition function of the
above-described model in the form \cite{Pat-Mryg-CM}:
\begin{eqnarray}
&&\Xi[\nu_{\alpha}]=\int \,({\rm d}\rho)({\rm d}\omega)\,
\exp\left(-\frac{\beta}{2V}\sum_{{\mathbf k}}[\widetilde
\phi^{SR}(k)\rho_{{\mathbf k},N}\rho_{-{\mathbf k},N} +\widetilde
\phi^{C}(k) \rho_{{\mathbf k},Q}\rho_{-{\mathbf k},Q}] \right.
\nonumber\\
&&\left. +{\rm
i}\sum_{{\mathbf k}}(\omega_{{\mathbf k},N}\rho_{{\mathbf
k},N}
+\omega_{{\mathbf k},Q}\rho_{{\mathbf k},Q})+\ln
\Xi_{HS}[\bar \nu_{N}-{\rm i}\omega_{N},\bar \nu_{Q}-{\rm
i}q_{\alpha}\omega_{Q}]
 \right).
 \label{Xi_full}
\end{eqnarray}
Here, the following notations are introduced. $\rho_{{\mathbf k},N}$
and $\rho_{{\mathbf k},Q}$ are the CVs which describe fluctuations
of the total number density and the charge density, respectively:
\begin{equation*}
\rho_{{\mathbf k},N}=\rho_{{\mathbf k},+}+\rho_{{\mathbf k},-},
\qquad \rho_{{\mathbf k},Q}=\rho_{{\mathbf k},+}-z\rho_{{\mathbf
k},-},
\end{equation*}
CV $\rho_{{\mathbf k},\alpha}=\rho_{{\mathbf k},\alpha}^c-{\rm
i}\rho_{{\mathbf k},\alpha}^s$ describes  the value of the $\mathbf
k$-th fluctuation mode of the number density of the $\alpha$th
species, the indices $c$ and $s$ denote real and imaginary parts of
$\rho_{{\mathbf k},\alpha}$; CVs $\omega_{N}$ and $\omega_{Q}$ are
conjugate to  $\rho_{N}$ and $\rho_{Q}$, respectively. $({\rm
d}\rho)$ and $({\rm d}\omega)$ denote  volume elements of the CV
phase space
\begin{displaymath}
({\rm d}\rho)=\prod_{A=N,Q}{\rm d}\rho_{0,A}{\prod_{\mathbf
k\not=0}}' {\rm d}\rho_{\mathbf k,A}^{c}{\rm d}\rho_{\mathbf
k,A}^{s}, \quad ({\rm d}\omega)=\prod_{A=N,Q}{\rm
d}\omega_{0,A}{\prod_{\mathbf k\not=0}}' {\rm d}\omega_{\mathbf
k,A}^{c}{\rm d}\omega_{\mathbf k,A}^{s}
\end{displaymath}
and the product over ${\mathbf k}$ is performed in the upper
semi-space ($\rho_{-\mathbf k,A}=\rho_{\mathbf k,A}^{*}$,
$\omega_{-\mathbf k,A}=\omega_{\mathbf k,A}^{*}$). Coefficients
$\widetilde{\phi}^{SR}(k)$ and $\widetilde{\phi}^{C}(k)$ are the
Fourier transforms of the corresponding interaction potentials. We
use the Weeks-Chandler-Andersen (WCA) regularization scheme
\cite{wcha,cha} for the both potentials, $\phi^{C}(r)$ and
$\phi^{SR}(r)$, inside the hard core.

$\Xi_{\text{HS}}[\bar \nu_{N}-{\rm i}\omega_{N},\bar \nu_{Q}-{\rm
i}q_{\alpha}\omega_{Q}]$ is the grand canonical partition function
of the hard-sphere system  with the renormalized chemical potentials
\begin{eqnarray*}
\bar\nu_{N}=\frac{z\bar\nu_{1}+\bar\nu_{2}}{1+z}, \qquad
\bar\nu_{Q}=\frac{\bar\nu_{1}-\bar\nu_{2}}{q_{0}(1+z)},
\end{eqnarray*}
where $\bar\nu_{\alpha}$ is determined by
\begin{eqnarray*}
\bar \nu_{\alpha}=\nu_{\alpha}+\frac{\beta}{2V}\sum_{{\mathbf
k}}[\widetilde\phi^{SR}(k)+q_{\alpha}^{2}\widetilde\phi^{C}(k)],
\end{eqnarray*}
$\nu_{\alpha}$ is the dimensionless chemical potential,
$\nu_{\alpha}=\beta\mu_{\alpha}-3\ln\Lambda_{\alpha}$ and
$\mu_{\alpha}$ is the chemical potential of the $\alpha$th species;
$\beta=1/k_{B}T$ is the reciprocal temperature;
$\Lambda_{\alpha}^{-1}=(2\pi m_{\alpha}\beta^{-1}/h^{2})^{1/2}$ is
the inverse de Broglie thermal wavelength. It is worth noting that
$\phi_{\alpha\beta}^{C}(r=0)$  is  a finite quantity due to the WCA
regularization. We introduce
\begin{equation}
{\rm i}\omega_{{\mathbf k},N}^{'}\delta_{{\mathbf
k}}={\rm i}\omega_{{\mathbf
k},N}\delta_{{\mathbf k}}-\Delta\nu_{N},
\label{nu_spheres}
\end{equation}
where $\Delta\nu_{N}=\bar\nu_{N}-\nu_{N, \text{HS}}$ with $\nu_{N, \text{HS}}$ being the
chemical potential of   hard spheres.

In order to develop the perturbation theory, we present
$\ln\Xi_{\text{HS}}[\nu_{N, \text{HS}}-{\rm i}\omega_{N}^{'},\bar \nu_{Q}-{\rm
i}q_{\alpha}\omega_{Q}]$ in the form of a cumulant expansion
\cite{patsahan-mryglod-patsahan:06}
\begin{eqnarray}
&&\ln\Xi_{\text{HS}}[\ldots]=\sum_{n\geq 0}\frac{(-{\rm
i})^{n}}{n!}\sum_{i_{n}\geq 0}
\sum_{{\mathbf{k}}_{1},\ldots,{\mathbf{k}}_{n}}
{\mathfrak{M}}_{n}^{(i_{n})}(k_{1},\ldots,k_{n})\omega_{{\bf{k}}_{1},Q}\ldots\omega_{{\bf{k}}_{i_{n}},Q}
\nonumber \\
&&\times \omega_{{\bf{k}}_{i_{n+1}},N}\ldots\omega_{{\bf{k}}_{n},N}
\delta_{{\bf{k}}_{1}+\ldots +{\bf{k}}_{n}},
 \label{2.11}
\end{eqnarray}
where the prime on $\omega_{{\bf{k}},N}$ is omitted for the sake of
simplicity. In Eq.~(\ref{2.11}), the $n$th cumulant
${\mathfrak{M}}_{n}^{(i_{n})}(k_{1},\ldots,k_{n})$ is   a linear
combination of the partial cumulants
${\mathfrak{M}}_{\alpha_{1}\ldots\alpha_{n}}(k_{1},\ldots,k_{n})$,
the superindex $i_{n}$ indicates the number of variables
$\omega_{{\bf{k}},Q}$ in the cumulant expansion. For details we
refer the reader to Ref.~\cite{Pat-Mryg-CM}.  The expressions for
${\mathfrak{M}}_{1}^{(i_{n})}$ and
 ${\mathfrak{M}}_{2}^{(i_{n})}$ are as follows:
\begin{eqnarray*}
{\mathfrak{M}}_{1}^{(0)}&=&\langle
N\rangle_{\text{HS}}, \qquad
{\mathfrak{M}}_{1}^{(1)}=0, \\ \nonumber
{\mathfrak{M}}_{2}^{(0)}(k)&=& {\widetilde G}_{2,{\text HS}}(k),
\qquad {\mathfrak{M}}_{2}^{(2)}(k)= q_{0}^{2}z\langle
N\rangle_{\text{HS}}\,\delta_{\mathbf k},
\end{eqnarray*}
where $\langle N\rangle_{\text{HS}}$ and ${\widetilde G}_{2,{\text
HS}}(k)$ are  the average  number of particles  and  the Fourier
transform of the two-particle connected correlation function of a
one-component hard-sphere system.
 It is worth noting that
${\mathfrak{M}}_{1}^{(1)}\equiv 0$ due to the electroneutrality
condition.
The recurrence formulas  for
${\mathfrak{M}}_{n}^{(i_{n})}$ are derived in Ref.~\cite{Pat-Mryg-CM} (see Eqs.~(46) in~\cite{Pat-Mryg-CM}).

Taking into account (\ref{nu_spheres})-(\ref{2.11}) and replacing
$\rho_{{\mathbf k},N}$ by $\rho_{{\mathbf
k},N}+{\mathfrak{M}}_{1}^{(0)}\delta_{{\mathbf k}}$  we can rewrite
Eq.~(\ref{Xi_full}) as follows:
\begin{eqnarray}
 &&
\Xi[\nu_{\alpha}]=\Xi_{\text{HS}}[\nu_{N, \text{HS}},\bar\nu_{Q}]\,{\cal{C}}
 \int \,({\rm d}\rho)({\rm d}\omega)\,\exp\left(\Delta\widetilde\nu_{N}\rho_{0,N}
-\frac{\beta}{2V}\sum_{{\mathbf k}}[\widetilde
\phi^{SR}(k)\rho_{{\mathbf k},N}\rho_{-{\mathbf k},N}
\right. \nonumber\\
&& \left. +\widetilde \phi^{C}(k)\rho_{{\mathbf k},Q}\rho_{-{\mathbf
k},Q}] +{\rm i}\sum_{{\mathbf k}}[\omega_{{\mathbf
k},N}\rho_{{\mathbf k},N}+\omega_{{\mathbf k},Q}\rho_{{\mathbf
k},Q}]+\sum_{n\geq 2}\frac{(-{\rm i})^{n}}{n!}\sum_{i_{n}\geq 0}
\sum_{{\mathbf{k}}_{1},\ldots,{\mathbf{k}}_{n}}
{\mathfrak{M}}_{n}^{(i_{n})}(k_{1},\ldots,k_{n})
\right. \nonumber\\
&& \left.
 +\omega_{{\bf{k}}_{1},Q}\ldots\omega_{{\bf{k}}_{i_{n}},Q}\,
\omega_{{\bf{k}}_{i_{n+1}},N}\ldots\omega_{{\bf{k}}_{n},N}
\delta_{{\bf{k}}_{1}+\ldots +{\bf{k}}_{n}}
 \right),
 \label{Xi_full_1}
\end{eqnarray}
where
\begin{eqnarray*}
&&\Delta\widetilde\nu_{N}=\Delta\nu_{N}-\beta\frac{\langle
N\rangle_{\text{HS}}}{V}\widetilde\phi^{SR}(0),
\\
&&{\cal{C}}=\exp\left[\Delta\nu_{N}\langle N\rangle_{\text{HS}}
 -\frac{\beta\langle N\rangle_{\text{HS}}^{2}}{2V}\widetilde\phi^{SR}(0)\right],
\end{eqnarray*}
and  $\ln\Xi_{\text{HS}}[\nu_{N, \text{HS}},\bar\nu_{Q}]={\mathfrak{M}}_{0}^{(0)}[\nu_{N, \text{HS}},\bar\nu_{Q}]$.
It is worth noting that  the Hamiltonian  in Eq.~(\ref{Xi_full_1})
does not include direct pair interactions of number density
fluctuations if $\widetilde{\phi}^{SR}(k)\equiv 0$.

Since we are interested in the gas-liquid critical point, the small-${\mathbf k}$ expansion of the cumulants can be considered.
Hereafter we will put
\begin{equation}
{\mathfrak{M}}_{2}^{(0)}(k)\simeq  {\mathfrak{M}}_{2}^{(0)}(0)+\displaystyle\frac{k^{2}}{2} {\mathfrak{M}}_{2,2}^{(0)}
\label{cumulant-2}
\end{equation}
and  approximate cumulants  for $n\geq 3$
by their values in the long-wavelength limit
\begin{equation}
{\mathfrak{M}}_{n}^{(i_{n})}(k_{1},\ldots,k_{n})\simeq
{\mathfrak{M}}_{n}^{(i_{n})}(0,\ldots).
\label{cumulant-n}
\end{equation}

\section{Effective Hamiltonian in the vicinity of the gas-liquid critical point}

In this section, based on Eqs.~(\ref{Xi_full_1})-(\ref{cumulant-n}),
we derive the effective Hamiltonian of the model~(\ref{2_1}) in the
vicinity of the gas-liquid critical point. We obtain consistently
all the coefficients, including the square-gradient term, within the
framework of the same approximation.

 First, we integrate   over CVs
$\omega_{{\bf{k}},N}$ and $\omega_{{\bf{k}},Q}$ in
Eq.~(\ref{Xi_full_1})  using the Gaussian density measure as the
basic one \cite{patsahan:04:1}. As a result, we get
\begin{eqnarray}
\Xi[\nu_{\alpha}]=\Xi_{\text{HS}}[\nu_{N, \text{HS}},\bar\nu_{Q}]\,{\cal{C}}\,
\prod_{{\mathbf k}}\frac{1}{\pi{\mathfrak{M}}_{2}^{(0)}}\prod_{{\mathbf k}}\frac{1}{\pi{\mathfrak{M}}_{2}^{(2)}}
 \int \,({\rm d}\rho)\, \exp\left(-{\cal{H}}[\nu_{\alpha},\rho_{N},\rho_{Q}]\right),
\label{Xi-rhoN-rhoQ}
\end{eqnarray}
where
\begin{eqnarray}
&&
-{\cal{H}}[\nu_{\alpha},\rho_{N},\rho_{Q}]= -a_{1}^{(0)}\rho_{0,N}-\frac{1}{2!}\sum_{\bf k}\left(a_{2}^{(0)}
\rho_{{\bf k},N}\rho_{-{\bf k},N}.+a_{2}^{(2)}\rho_{{\bf k},Q}\rho_{-{\bf k},Q}\right)
\nonumber \\
&&
-\frac{1}{3!}\sum_{{\mathbf{k}}_{1},{\mathbf{k}}_{2},{\mathbf{k}}_{3}}\left(a_{3}^{(0)}\rho_{{\bf
k_{1}},N} \rho_{{\bf k_{2}},N}\rho_{{\bf
k_{3}},N}+3a_{3}^{(2)}\rho_{{\bf k_{1}},N}\rho_{{\bf
k_{2}},Q}\rho_{{\bf k_{3}},Q}+a_{3}^{(3)}\rho_{{\bf k_{1}},Q}\rho_{{\bf
k_{2}},Q}\rho_{{\bf k_{3}},Q}
\right)
\nonumber \\
&&\times\delta_{{\mathbf{k}}_{1}+{\mathbf{k}}_{2}+{\mathbf{k}}_{3}}
-\frac{1}{4!}\sum_{{\mathbf{k}}_{1},\ldots,{\mathbf{k}}_{4}}\left(a_{4}^{(0)}\rho_{{\bf
k_{1}},N} \ldots\rho_{{\bf k_{4}},N}+6a_{4}^{(2)}\rho_{{\bf k_{1}},N} \rho_{{\bf
k_{2}},N}\rho_{{\bf k_{3}},Q}\rho_{{\bf k_{4}},Q} \right.
\nonumber \\
&&\left.
+4a_{4}^{(3)}\rho_{{\bf k_{1}},N} \rho_{{\bf k_{2}},Q}\rho_{{\bf
k_{3}},Q}\rho_{{\bf k_{4}},Q}
+a_{4}^{(4)}\rho_{{\bf k_{1}},Q} \rho_{{\bf k_{2}},Q}\rho_{{\bf
k_{3}},Q}\rho_{{\bf k_{4}},Q}\right)
\delta_{{\mathbf{k}}_{1}+\ldots+{\mathbf{k}}_{4}} +\ldots
\label{H-eff1}
\end{eqnarray}
and the coefficients $a_{n}^{(i_{n})}$ ($n\leq 4$)  are given by
\begin{equation}
a_{1}^{(0)}=-\Delta\widetilde\nu_{N}, \qquad
a_{1}^{(1)}=0,
\label{a1_0}
\end{equation}
\begin{equation}
a_{2}^{(0)}=\frac{\beta}{V}\widetilde{\phi}^ {SR}(k)+\frac{1}{{\mathfrak{M}}_{2}^{(0)}(k)}, \qquad
a_{2}^{(2)}=\frac{\beta}{V}\widetilde{\phi}^ {C}(k)+\frac{1}{{\mathfrak{M}}_{2}^{(2)}},
\label{a2_0}
\end{equation}
\begin{equation}
a_{3}^{(0)}=-\frac{{\mathfrak{M}}_{3}^{(0)}}{({\mathfrak{M}}_{2}^{(0)})^{3}},
\qquad
a_{3}^{(2)}=-\frac{{\mathfrak{M}}_{3}^{(2)}}{{\mathfrak{M}}_{2}^{(0)}
({\mathfrak{M}}_{2}^{(2)})^{2}},
\qquad
a_{3}^{(3)}=-\frac{{\mathfrak{M}}_{3}^{(3)}}{({\mathfrak{M}}_{2}^{(2)})^{3}},
\label{a3_0}
\end{equation}
\begin{eqnarray}
a_{4}^{(0)}&=&-\frac{1}{({\mathfrak{M}}_{2}^{(0)})^{4}}\left({\mathfrak{M}}_{4}^{(0)}
-3\frac{({\mathfrak{M}}_{3}^{(0)})^{2}}{{\mathfrak{M}}_{2}^{(0)}}\right),
\nonumber \\
a_{4}^{(2)}&=&-\frac{1}{({\mathfrak{M}}_{2}^{(0)})^{2}
({\mathfrak{M}}_{2}^{(2)})^{2}}\left({\mathfrak{M}}_{4}^{(2)}-\frac{{\mathfrak{M}}_{3}^{(0)}{\mathfrak{M}}_{3}^{(2)}}{{\mathfrak{M}}_{2}^{(0)}}-
2\frac{({\mathfrak{M}}_{3}^{(2)})^{2}}{{\mathfrak{M}}_{2}^{(2)}} \right),
\nonumber \\
a_{4}^{(3)}&=&-\frac{1}{{\mathfrak{M}}_{2}^{(0)}({\mathfrak{M}}_{2}^{(2)})^{3}}\left({\mathfrak{M}}_{4}^{(3)}-
3\frac{{\mathfrak{M}}_{3}^{(2)}{\mathfrak{M}}_{3}^{(3)}}{{\mathfrak{M}}_{2}^{(2)}}\right),
\nonumber \\
a_{4}^{(4)}&=&-\frac{1}{({\mathfrak{M}}_{2}^{(2)})^{4}}\left({\mathfrak{M}}_{4}^{(4)}
-3\frac{({\mathfrak{M}}_{3}^{(2)})^{2}}{{\mathfrak{M}}_{2}^{(0)}}
-3\frac{({\mathfrak{M}}_{3}^{(3)})^{2}}{{\mathfrak{M}}_{2}^{(2)}}\right).
\label{a4_0}
\end{eqnarray}
A distinguishing feature of the
Hamiltonian(\ref{H-eff1})-(\ref{a4_0}), unlike that for the RPM
\cite{patsahan:04:1}, is the presence of the coefficients
$a_{n}^{(3)}$ due to the charge-asymmetry of the model.
Using the recurrence formulae from Ref.~\cite{Pat-Mryg-CM},
Eqs.~(\ref{a1_0})-(\ref{a4_0}) can be reduced to the form given in
Appendix~A.

\subsection{Gaussian approximation}

Let us consider the Gaussian approximation, which corresponds to
Eq.~(\ref{H-eff1}) where only the terms with $n\leq 2$ are taken
into account. In this case, the integration over CVs  $\rho_{{\bf
k},N}$ and $\rho_{{\bf k},Q}$ in Eq.~(\ref{Xi-rhoN-rhoQ}) leads to
the following expression for the logarithm  of the grand partition
function
\begin{eqnarray*}
&&\ln\Xi_{\text{G}}=\ln \Xi_{\mathrm{HS}}+\frac{\beta\langle
N\rangle_{\text{HS}}^{2}}{2V}\widetilde{\phi}^{\text{SR}}(0)
-\frac{1}{2}\sum_{\mathbf{k}}\ln\left[1+\frac{\beta}{V}\widetilde{\phi}^{\text{SR}}(k)
{\mathfrak{M}}_{2}^{(0)}(k)\right] \nonumber \\
&&-\frac{1}{2}\sum_{\mathbf{k}}\ln\left[1+\frac{\beta}{V}\widetilde{\phi}^{C}(k)
{\mathfrak{M}}_{2}^{(2)}\right].
\end{eqnarray*}
The Legendre transform of $\ln\Xi_{G}$ yields  the Helmholtz free
energy in the one-loop approximation. The result is
\begin{eqnarray}
&&\beta f_{\text{RPA}}=\frac{\beta F_{\text{RPA}}}{V}=\beta
f_{\text{HS}}-\frac{\beta\rho}{2V}\sum_{\mathbf{k}}\widetilde{\phi}^{\text{SR}}(k)-\frac{\beta
q_{0}^{2}z\rho}{2V}\sum_{\mathbf{k}}\widetilde{\phi}^{\text{C}}(k)
+\frac{1}{2}\beta\rho^{2}\widetilde{\phi}^{\text{SR}}(0)
\nonumber\\
&&+\frac{1}{2V}\sum_{\mathbf
k}\ln\left[1+\beta\rho\widetilde{\phi}^{\text{SR}}S_{2,\text{HS}}(k)\right]+\frac{1}{2V}\sum_{\mathbf
k}\ln\left[1+\beta q_{0}^{2}z\rho\widetilde{\phi}^{C}(k)\right], \label{da2.10a}
\end{eqnarray}
where $f_{\ldots}$ denotes the Helmholtz free energy density. Again,
the subscript $\text{HS}$ refers to the hard-sphere system,
$S_{2,\text{HS}}(k)$ is the pair structure factor of a one-component
hard-sphere system \cite{hansen_mcdonald}. The one-loop free energy
(\ref{da2.10a}) coincides with the free energy in the random phase
approximation (RPA) of the theory of liquids \cite{hansen_mcdonald}.

From  (\ref{da2.10a}) one can find the chemical potential $\nu_{N}$
in the RPA
\begin{eqnarray}
\nu_{N}^{\text{RPA}}&=&\nu_{N,\text{HS}}-\frac{\beta}{2V}\sum_{\mathbf{k}}\widetilde{\phi}^{\text{SR}}(k)
-\frac{\beta
q_{0}^{2}z}{2V}\sum_{\mathbf{k}}\widetilde{\phi}^{\text{C}}(k)
+\beta\rho\widetilde{\phi}^{\text{SR}}(0)
\nonumber \\
&&+\frac{1}{2V}\sum_{\mathbf
k}\frac{\beta q_{0}^{2}z\widetilde{\phi}^{C}(k)}{1+\beta q_{0}^{2}z\rho\widetilde{\phi}^{C}(k)}.
\label{nu_N_rpa}
\end{eqnarray}
and  obtain on its basis the mean-field gas-liquid phase diagram.
The gas-liquid critical parameters of the charge-asymmetric PM
($\phi^{\text{SR}}(r)=0$)   were calculated in Ref.~\cite{Cai-Mol1}
using  the above equations and different regularization schemes for
the Coulomb potential inside the hard core.

\subsection{Coefficients of the Landau-Ginzburg-Wilson Hamiltonian}

Our aim is to derive the effective Hamiltonian in terms  of CVs
$\rho_{{\bf k},N}$ related to the order parameter associated with
the gas-liquid critical point. To this end, we integrate out  CVs
$\rho_{{\bf k},Q}$ in Eq.~(\ref{H-eff1}) following the programme
outlined for the  RPM \cite{patsahan:04:1}. As a result, we arrive
at the effective LGW  Hamiltonian of the following form:
\begin{eqnarray}
-{\cal H}^{eff}[\rho_{N}]=-\sum_{n\geq 1}\frac{1}{n!\langle
N\rangle^{n-1}}\sum_{{\mathbf{k}}_{1},\ldots,{\mathbf{k}}_{n}}
a_{n}\;\rho_{{\bf k_{1}},N}\, \rho_{{\bf k_{2}},N}\,\ldots\rho_{{\bf
k_{n}},N}\, \delta_{{\mathbf{k}}_{1}+\ldots+{\mathbf{k}}_{n}}.
\label{H-rho}
\end{eqnarray}
Coefficient $a_{n}$ can be presented as
\begin{equation}
a_{n}=a_{n}^{(0)}+\Delta a_{n}, \label{dA.19a}
\end{equation}
where  the second addend is the correction obtained after
integration over CVs $\rho_{{\bf k},Q}$. Each correction $\Delta
a_{n}$ has the form of infinite  series.  In particular,  after some
algebra we find the following expressions for $\Delta a_{n}$ in the
approximation corresponding to a   one sum over the wave vector
$\mathbf{q}$
\begin{eqnarray}
\Delta a_{1}=-\frac{1}{2<N>}\sum_{\mathbf{q}}\widetilde
G_{QQ}(q)+\ldots, \label{a1}
\end{eqnarray}
\begin{eqnarray}
\Delta a_{2}(k)=\frac{1}{<N>}\sum_{\mathbf{q}}\widetilde
G_{QQ}(q)-\frac{1}{2<N>}\sum_{\mathbf{q}}\widetilde
G_{QQ}(q)\widetilde G_{QQ}(|\mathbf{q}+\mathbf{k}|) + \ldots,
\label{a2}
\end{eqnarray}
\begin{eqnarray}
&&\Delta
a_{3}(k_{1},k_{2})=-\frac{3}{<N>}\sum_{\mathbf{q}}\widetilde
G_{QQ}(q)+\frac{3}{<N>}\sum_{\mathbf{q}}\widetilde
G_{QQ}(q)\widetilde
G_{QQ}(|\mathbf{q}+\mathbf{k_{1}}|) \nonumber\\
&&-\frac{1}{<N>}\sum_{\mathbf{q}}\widetilde G_{QQ}(q)\widetilde
G_{QQ}(|\mathbf{q}+\mathbf{k_{1}}|) \widetilde
G_{QQ}(|\mathbf{q}-\mathbf{k_{2}}|) + \ldots, \label{a3}
\end{eqnarray}
\begin{eqnarray}
&&\Delta
a_{4}(k_{1},k_{2},k_{3})=\frac{12}{<N>}\sum_{\mathbf{q}}\widetilde
G_{QQ}(q)-\frac{6}{<N>}\sum_{\mathbf{q}}\widetilde
G_{QQ}(q)\widetilde
G_{QQ}(|\mathbf{q}+\mathbf{k_{1}}+\mathbf{k_{2}}|) \nonumber\\
&&-\frac{12}{<N>}\sum_{\mathbf{q}}\widetilde G_{QQ}(q)\widetilde
G_{QQ}(|\mathbf{q}+\mathbf{k_{1}}|)+\frac{12}{<N>}\sum_{\mathbf{q}}\widetilde
G_{QQ}(q)\widetilde G_{QQ}(|\mathbf{q}+\mathbf{k_{1}}|)
\widetilde G_{QQ}(|\mathbf{q}-\mathbf{k_{2}}-\mathbf{k_{3}}|)\nonumber\\
&&-\frac{3}{<N>}\sum_{\mathbf{q}}\widetilde G_{QQ}(q)\widetilde
G_{QQ}(|\mathbf{q}+\mathbf{k_{1}}|) \widetilde
G_{QQ}(|\mathbf{q}-\mathbf{k_{2}}|)\widetilde
G_{QQ}(|\mathbf{q}-\mathbf{k_{2}}-\mathbf{k_{3}}|) +\ldots,
\label{a4}
\end{eqnarray}
where we have made use of the formulas from Appendix~A. Hereafter,
for notation simplicity,  the subscript $\text{HS}$ is omitted.

$\widetilde G_{QQ}(q)$ is the Fourier transform of a charge-charge
connected correlation function of the charge-asymmetric PM
determined in the Gaussian approximation
\begin{equation*}
\widetilde G_{QQ}(q)=\frac{1}{1+\beta\rho q_{0}^{2}z\widetilde\phi^{C}(q)}.
\end{equation*}
For the WCA regularization \cite{cha},
\begin{equation}
\widetilde\phi^{C}(q)=4\pi\displaystyle\frac{\sin q\sigma}{\epsilon
q^{3}}. \label{wca_coul}
\end{equation}
Now, we are in a   position to study   Eqs.~(\ref{a1})-(\ref{a4}) in
detail. As is seen from Eqs~(\ref{a1})-(\ref{a4}),  corrections
$\Delta a_{n}$ for $n\geq 2$ depend  on the wave vectors. Expanding
the second term of $\Delta a_{2}$ at small $k$ one can readily see
that  the linear term  vanishes. As a result,   $\Delta a_{2}$  can
be presented as follows:
\begin{equation}
\Delta a_{2}=\Delta a_{2,0}+k^{2}\Delta a_{2,2}+ \mathcal{O}(k^{4}),
\label{dA.22}
\end{equation}
where
\begin{eqnarray}
\Delta a_{2,0}&=&\frac{1}{\langle N\rangle}
\sum_{{\mathbf{q}}}\widetilde G_{QQ}(q)-\frac{1}{2\langle N\rangle}
\sum_{{\mathbf{q}}}[\widetilde G_{QQ}(q)]^{2}, \nonumber \\
\Delta a_{2,2}&=-&\frac{1}{4\langle
N\rangle}\sum_{\mathbf{q}}\widetilde G_{QQ}(q)\widetilde
G_{QQ}^{(2)}(q) \label{dA.23}
\end{eqnarray}
and the superscript $(2)$ refers to  the second-order derivative
with respect to the wave vector $k$: $\widetilde
G_{QQ}^{(2)}(q)=\partial^{2}\widetilde
G_{Q^{}Q}(|\mathbf{q}+\mathbf{k}|)/
\partial k^{2}\rvert_{k=0}$.
The appearance of the factor  $k^{2}$   in Eq.~(\ref{dA.22}) is
caused by the charge-charge correlations being  taken into account.
It should be noted that the expression for $\Delta a_{2,2}$ has the
same form as that obtained  for the RPM in
Ref.~\cite{patsahan:04:1}. However, the numerical estimates of the
effective Hamiltonian coefficients were not  made in that work. It
will be done below.

The integrals   $\int\left[\widetilde
G_{QQ}(q)\right]^{n}{\rm d}\mathbf{q}$ entering Eqs.~(\ref{a1})-(\ref{a4}) 
are divergent at  $q\rightarrow\infty$. This divergence can be avoided
by introducing the cutoff wave-vector which, however,  leads to the
cutoff-dependent results.  Another way is to rearrange the terms in
Eqs.~(\ref{a1})-(\ref{a4}) expressed by infinite series. Here, we
will follow the second way.

We    approximate $\Delta a_{2}$  by
Eqs.~(\ref{dA.22})-(\ref{dA.23}) and replace $\Delta
a_{3}(k_{1},\ldots)$ and $\Delta a_{4}(k_{1},\ldots)$ by their
values in the long-wavelength limit  putting $\Delta a_{3}(k_{1},\ldots)\simeq \Delta
a_{3,0}$ and $\Delta a_{4}(k_{1},\ldots)\simeq \Delta a_{4,0}$. It is
convenient  to calculate $\Delta a_{n}$ using  Feynman diagram
presentation. Limiting the series in  Eqs.~(\ref{a1})-(\ref{a4}) to the order  of one sum over
$\mathbf{q}$
one can present  the
part of the effective Hamiltonian (\ref{H-rho})  in the following
diagrammatic form
\begin{eqnarray}
\Delta a_{1}\rho_{0,N}=-\frac{1}{2}\raisebox{-0.4\height}%
{\includegraphics[height=8mm]{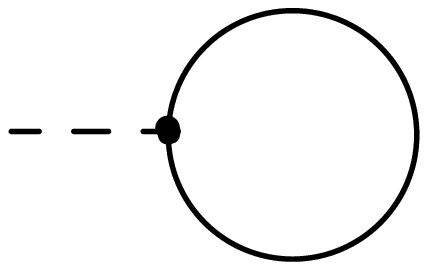}}, \label{a1_diag}
\end{eqnarray}
\begin{eqnarray}
\frac{1}{2!\langle N\rangle}\sum_{{\mathbf{k}}}\Delta a_{2,0}\rho_{{\bf k},N}\rho_{-{\bf k},N}= \raisebox{-0.4\height}%
{\includegraphics[height=8mm]{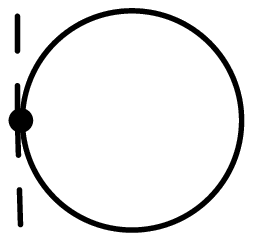}}-\frac{1}{2}\,\raisebox{-0.4\height}%
{\includegraphics[height=8mm]{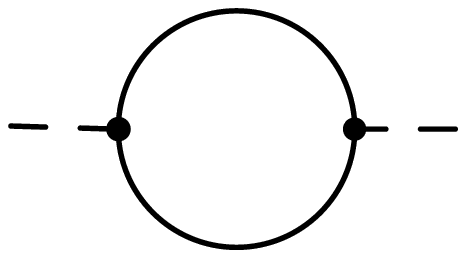}},  \label{a2_diag}
\end{eqnarray}
\begin{eqnarray}
\frac{1}{3!\langle
N\rangle^{2}}\sum_{{\mathbf{k}}_{1},{\mathbf{k}}_{2}}\Delta a_{3,0}
\rho_{{\bf k_{1}},N}\rho_{{\bf k_{2}},N}\rho_{-{\bf k_{1}}-{\bf k_{2}},N}=-3\,\raisebox{-0.4\height}%
{\includegraphics[height=8mm]{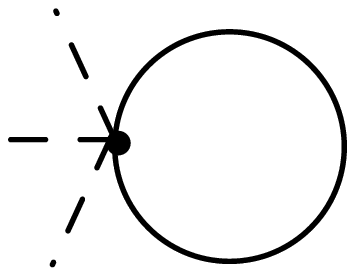}}+3\,\raisebox{-0.4\height}%
{\includegraphics[height=8mm]{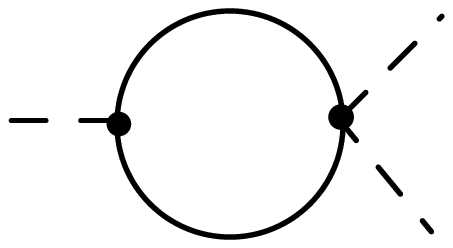}}-\raisebox{-0.32\height}%
{\includegraphics[height=12mm]{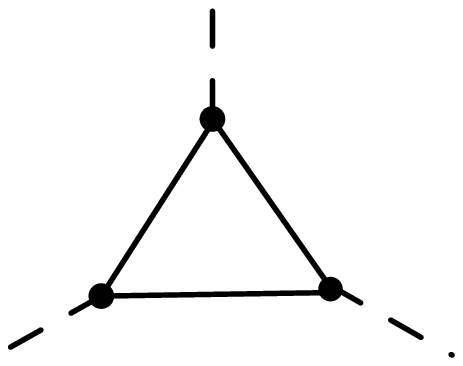}},   \label{a3_diag}
\end{eqnarray}
\begin{eqnarray}
&&\frac{1}{4!\langle
N\rangle^{3}}\sum_{{\mathbf{k}}_{1},{\mathbf{k}}_{2},{\mathbf{k}}_{3}}\Delta
a_{4,0}\rho_{{\bf k_{1}},N}\rho_{{\bf k_{2}},N}\rho_{{\bf
k_{3}},N}\rho_{-{\bf k_{1}}-{\bf k_{2}}-{\bf k_{3}},N}=12\,\raisebox{-0.4\height}%
{\includegraphics[height=8mm]{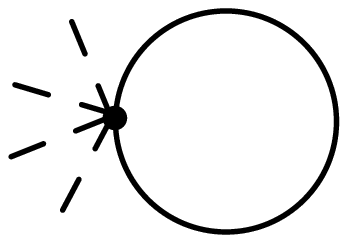}}-6\,\raisebox{-0.4\height}%
{\includegraphics[height=8mm]{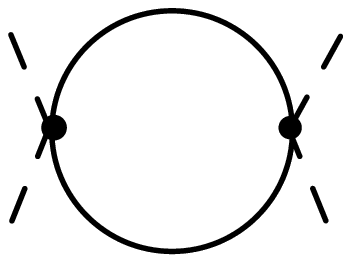}}\nonumber
\\ &&
-12\,\raisebox{-0.4\height}%
{\includegraphics[height=8mm]{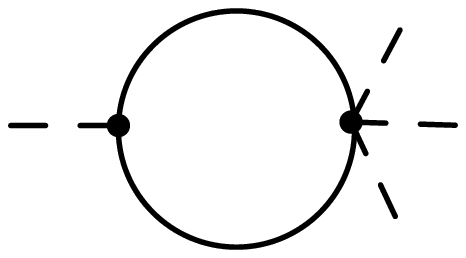}}
+12\raisebox{-0.35\height}%
{\includegraphics[height=10mm]{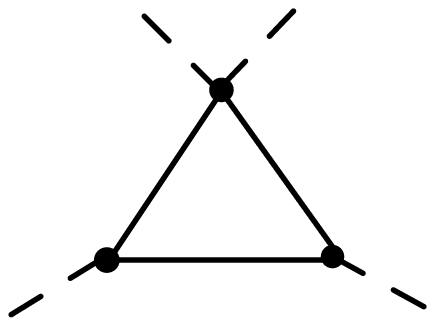}}-3\raisebox{-0.35\height}%
{\includegraphics[height=10mm]{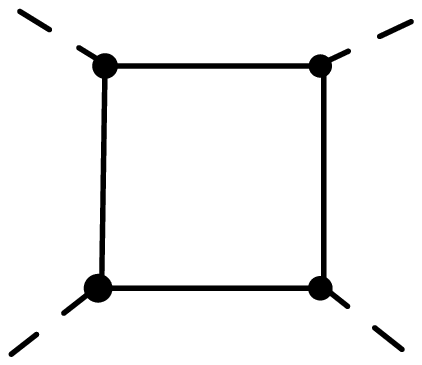}}.
 \label{a4_diag}
\end{eqnarray}
In Eqs.~(\ref{a1_diag})-(\ref{a4_diag}), each external leg
represents CV $\rho_{{\bf k},N}$, the vertex represents the factor
$\displaystyle\frac{a_{n}^{(i_{n})}}{n!}\delta_{{\mathbf{k}}_{1}+\ldots+{\mathbf{k}}_{n}}$
and each internal line corresponds to the propagator $\widetilde
G_{QQ}(q)$. Integrals over  wave vectors are implied here. As is
seen, the restriction   to a one sum over $\mathbf{q}$ in the
expressions for $\Delta a_{n}$ leads to the one-loop diagrams having
the same number of vertices and bonds.

Now, introducing
\begin{equation*}
\widetilde g(q)=-\frac{\beta\rho q_{0}^{2}z\widetilde\phi^{C}(q)}{1+\beta\rho
q_{0}^{2}z\widetilde\phi^{C}(q)},
\end{equation*}
we substitute
\begin{equation}
\widetilde G_{QQ}(q)=1+\widetilde g(q) \label{G-g}
\end{equation}
into Eqs.~(\ref{a1})-(\ref{a4}). Then,   adding all the terms
corresponding to the one-loop diagrams in
Eqs~(\ref{a1_diag})-(\ref{a4_diag}) we  present $\Delta a_{n}$ in
the form:
\begin{eqnarray}
\Delta a_{1}=-\frac{1}{2<N>}\sum_{\mathbf{q}}\widetilde g(q),
\label{a1_g}
\end{eqnarray}
\begin{eqnarray}
\Delta a_{2}(k)=-\frac{1}{2<N>}\sum_{\mathbf{q}}\left[\widetilde
g(q)\right]^{2}-\frac{k^{2}}{4 \langle
N\rangle}\sum_{\mathbf{q}}\widetilde g^{(2)}(q)\left[1+\widetilde
g(q)\right], \label{a2_g}
\end{eqnarray}
\begin{eqnarray}
\Delta a_{3}=-\frac{1}{<N>}\sum_{\mathbf{q}}\left[\widetilde
g(q)\right]^{3}, \label{a3_g}
\end{eqnarray}
\begin{eqnarray}
\Delta a_{4}=-\frac{3}{<N>}\sum_{\mathbf{q}}\left[\widetilde
g(q)\right]^{4}. \label{a4_g}
\end{eqnarray}
The integrals entering Eqs.~(\ref{a1_g})-(\ref{a4_g}) are
convergent in the lower and upper limits. Again, the superscript $(2)$ denotes the second-order
derivative with respect to $k$. Comparing Eqs.~(\ref{a2_g}) and
(\ref{dA.23}) it is easy to check  that substitution (\ref{G-g})
does not change the  coefficient $\Delta a_{2,2}$. Finally,
inserting Eqs.~(\ref{a1_g})-(\ref{a4_g}) in (\ref{dA.19a}) we arrive
at explicit expressions for the coefficients $a_{n}$ in the one-loop
approximation:
\begin{eqnarray}
&& a_{1,0}=-\Delta\widetilde\nu_{N}-\widetilde{\cal C}_{1,\text{C}},
\label{a1_rpa}\\
&& a_{2,0}=-\rho\,\widetilde {\cal
C}_{2,\text{HS}}+\beta\rho\widetilde\phi^{SR}(0)-\rho\,\widetilde
{\cal C}_{2,\text{C}},
\label{a20_rpa}\\
&& a_{2,2}=-\frac{1}{2}\rho\,\widetilde {\cal
C}_{2,\text{HS}}^{(2)}+\frac{1}{2}\beta\rho\widetilde\phi^{SR,(2)}
-\frac{1}{4 \langle N\rangle}\sum_{\mathbf{q}}\widetilde
g^{(2)}(q)\left[1+\widetilde g(q)\right],
\label{a22_rpa}\\
&& a_{3,0}=-\rho^{2}\widetilde{\cal
C}_{3,\text{HS}}-\rho^{2}\widetilde{\cal C}_{3, \text{C}},
\label{a3_rpa}\\
&& a_{4,0}=-\rho^{3}\widetilde{\cal
C}_{4,\text{HS}}-\rho^{3}\widetilde{\cal C}_{4, \text{C}},
\label{a4_rpa}
\end{eqnarray}
where  $a_{n,0}=a_{n}|_{k_{i}=0}$,  $a_{2,0}$  and $a_{2,2}$ are the
coefficients of the expansion $a_{2}=a_{2,0}+k^{2}a_{2,2}$, and
$\widetilde {\cal C}_{2,\text{HS}}(k)=\widetilde {\cal
C}_{2,\text{HS}}+\frac{1}{2}k^{2}\widetilde {\cal
C}_{2,\text{HS}}^{(2)}$. $\widetilde{\cal C}_{n,\text{HS}}$ is the
Fourier transform of the $n$-particle direct correlation function of
a one-component hard-sphere system in the long-wavelength limit
\begin{eqnarray}
-\rho\,\widetilde{\cal
C}_{2,\text{HS}}=\frac{1}{S_{2,\text{HS}}(0)}, \qquad
\rho^{n-1}\widetilde{\cal C}_{n,\text{HS}}=\rho^{n-1}\widetilde
{\cal C}_{n,\text{HS}}(0,\ldots)=a_{n}^{(0)}, \qquad n\geq 3
\end{eqnarray}
and $\widetilde {\cal C}_{2,\text{HS}}^{(2)}=\partial^{2} \widetilde
{\cal C}_{2,\text{HS}}(k)/\partial k^{2}|_{k=0}$. Using the formulas
from Appendix~A, one can establish a link between the direct
correlation functions $\widetilde{\cal C}_{n,\text{HS}}$ and the
connected correlation functions $\widetilde G_{n,\text{HS}}$ for
$n\leq 4$.

$\widetilde{\cal C}_{n,\text{C}}$ denotes the contribution to the
$n$-particle  direct correlation function from the charge subsystem
\begin{equation}
\rho^{n-1}\widetilde{\cal C}_{n, \text{C}}=\rho^{n-1}\widetilde{\cal
C}_{n, \text{C}}(k=0)=\frac{(n-1)!}{2}\frac{1}{\langle
N\rangle}\sum_{\mathbf{q}}\left[\widetilde g(q)\right]^{n}.
\label{Cn_C}
\end{equation}
It is easy to check that  the coefficients $a_{n,0}$  for $n\geq 2$
can be obtained from the one-loop free energy (see
Eq.~(\ref{da2.10a}))
\begin{eqnarray*}
 a_{n,0}=\rho^{n-1}\frac{\partial^{n}(-\beta f_{\text{RPA}})}{\partial
 \rho^{n}}=\rho^{n-1}\widetilde{\cal C}_{n}(0, \ldots),
\end{eqnarray*}
where $\widetilde{\cal C}_{n}(0, \ldots)$ denotes the Fourier
transform of the n-particle  direct correlation function of the full
system in the long-wavelength limit. It is worth noting that the
functions $\widetilde{\cal C}_{n}$ differ from the ordinary direct
correlation functions $\widetilde c_{n}$ by an ideal term
\cite{hansen_mcdonald,Stell:75}.

Coefficient $ a_{1,0}$ is the
excess part of the chemical potential $\nu_{N}$ connected with the
short-range attractive and long-range Coulomb interactions. Equation
$a_{1,0}=0$ leads to the expression for $\nu_{N}$ in the RPA (see
Eq.~(\ref{nu_N_rpa})). Coefficient $a_{2,2}$ describes the square-gradient
term. We emphasize that all the  coefficients given by
Eqs~(\ref{a1_rpa})-(\ref{a4_rpa}) are found in the one-loop
approximation corresponding to a one sum over the wave vector. The charge asymmetry
does not manifest itself in the above
equations and, therefore, there is no difference between the RPM and
the charge-asymmetric PM at this level of approximation. The
resulting effective Hamiltonian  reads as
\begin{eqnarray*}
&&{\cal H}^{eff}=a_{1,0}\rho_{0,N}+\frac{1}{2!\langle
N\rangle}\sum_{{\mathbf{k}}}\left(a_{2,0}+k^{2}a_{2,2}\right)\rho_{{\bf
k},N}\rho_{-{\bf k},N}+\frac{1}{3!\langle
N\rangle^{2}}\sum_{{\mathbf{k}}_{1},{\mathbf{k}}_{2}} a_{3,0}
\nonumber \\
&& \times\rho_{{\bf k_{1}},N}\rho_{{\bf k_{2}},N}\rho_{-{\bf
k_{1}}-{\bf k_{2}},N}+\frac{1}{4!\langle
N\rangle^{3}}\sum_{{\mathbf{k}}_{1},{\mathbf{k}}_{2},{\mathbf{k}}_{3}}a_{4,0}\rho_{{\bf
k_{1}},N}\rho_{{\bf k_{2}},N}\rho_{{\bf k_{3}},N}\rho_{-{\bf
k_{1}}-{\bf k_{2}}-{\bf k_{3}},N}.
\end{eqnarray*}
Therefore, within the framework of the same approximation, we have
derived the microscopic-based expressions for the coefficients of
the effective Hamiltonian. Taking into account the charge-charge
correlations through integration over the charge subsystem (the CVs
$\rho_{{\bf k},Q}$) we get a contribution to the coefficients at the
second order which describes the effective attraction of short-range
character. The resulting Hamiltonian has the structure of the LGW
Hamiltonian of an Ising model in an external magnetic field.
Eqs.~(\ref{a1_rpa})-(\ref{Cn_C}) will be used below for the
calculation of the Ginzburg temperature.

\section{Ginzburg temperature}

We present $a_{2,0}$ in the form:
\begin{eqnarray*}
a_{2,0}=\bar a_{2,0}+a_{2,t}\,t, \qquad t=\frac{T-T_{c}}{T_{c}},
\end{eqnarray*}
where $a_{2,0}=\bar a_{2,0}(t=0)$ and $a_{2,t}=\left.\displaystyle\frac{\partial
a_{2,0}}{\partial t}\right|_{t=0}$. Hereafter, the subscript $c$ refers to
the critical value.

At the critical point, the  system of equations
 \begin{eqnarray}
  \bar a_{2,0}=0, \qquad a_{3,0}=0
\label{cr-point}
 \end{eqnarray}
 holds  yielding the critical temperature and the critical density.
 After the substitution of $T_{c}$ and  $\rho_{c}$ in the equation $ a_{1,0}=0$ one gets the
 critical value of the chemical potential $\nu_{N}$.

 Following \cite{Goldenfeld,fisher3},   the reduced Ginzburg temperature can be
written  as follows:
\begin{eqnarray}
 t_{G}=\displaystyle\frac{1}{32\pi^{2}}\frac{a_{4,0}^{2}}{a_{2,\tau} a_{2,2}^{3}},
\label{t_G}
\end{eqnarray}
where all the  coefficients should be calculated at the critical
temperature and the critical density determined from
Eqs.~(\ref{cr-point}).

\subsection{Hard-core square-well model}

First, we consider a one-component system of hard spheres
interacting through the SW potential of depth $\varepsilon$ and
range $\lambda=1.5\sigma$. This system exhibits a typical Ising
critical behaviour. For the WCA regularization, the Fourier
transform of the SW potential has the form \cite{wcha}
\begin{equation}
\widetilde \phi^{SR}(k)=\widetilde\phi^{SR}(0)\frac{3}{(\lambda x)^3}[-\lambda x
~\cos(\lambda x) + \sin(\lambda x)],
\label{phi-sr-k}
\end{equation}
where
$x=k\sigma$ and $\widetilde \phi^{SR}(0)= -\varepsilon\sigma^3 \frac{4\pi}{3} \lambda^3$.

For this model $\Delta a_{n}\equiv 0$,  and we get from
Eqs.~(\ref{a1_rpa})-(\ref{a4_rpa})  and Eq.~(\ref{phi-sr-k}) simple
expressions for the coefficients of the LGW Hamiltonian
\begin{eqnarray}
a_{1,0}&=&-\Delta\nu_{N}+\frac{1}{2T^{\text{SR}}} -\frac{8\eta\lambda^{3}}{T^{\text{SR}}}, \label{a1_sw} \\
\bar a_{2,0}&=&-\rho\,\widetilde {\cal
C}_{2,\text{HS}}-\frac{8\eta\lambda^{3}}{T^{\text{SR}}_{c}}, \qquad
a_{2,t}=\frac{8\eta\lambda^{3}}{T^{\text{SR}}_{c}}, \qquad
a_{2,2}=-\frac{1}{2}\rho\,\widetilde
{\cal C}_{2,\text{HS}}^{(2)}+\frac{4}{5}\frac{\eta\lambda^{5}}{T^{\text{SR}}}, \label{a22_sw}\\
a_{3,0}&=&-\rho^{2}\widetilde {\cal C}_{3,\text{HS}}, \qquad
a_{4,0}=-\rho^{3}\widetilde {\cal C}_{4,\text{HS}}, \label{a3-a4_sw}
\end{eqnarray}
where $\Delta\nu_{N}=\nu_{N}-\nu_{N,\text{HS}}$. The reduced
temperature in the hard-core square-well (HCSW) model,
$T^{\text{SR}}$, is defined as the ratio between the thermal energy
and the  interaction strength of the two hard spheres at  contact
$T^{\text{SR}}=k_{\text{B}}T/\varepsilon$,  and
$\eta=\pi\rho\sigma^{3}/6$ is the packing fraction.  We use the
Percus-Yevick (PY) approximation for the $n$-particle direct
correlation functions of the hard-sphere system
\cite{hansen_mcdonald,ashcroft-1}. As a result, we have
\begin{eqnarray*}
-\rho\,\widetilde {\cal
C}_{2,\text{HS}}=\frac{(1+2\eta)^{2}}{(1-\eta)^{4}}, \qquad
-\rho\,\widetilde {\cal
C}_{2,\text{HS}}^{(2)}=\frac{\eta(16-11\eta+4\eta^{2})}{10(1-\eta)^{4}},
\end{eqnarray*}
\begin{equation*}
-\rho^{2}\widetilde {\cal
C}_{3,\text{HS}}=\frac{(1-7\eta-6\eta^{2})(1+2\eta)}{(1-\eta)^{5}},
\end{equation*}
\begin{equation*}
-\rho^{3}\widetilde {\cal
C}_{4,\text{HS}}=\frac{2(1-6\eta+15\eta^{2}+56\eta^{3}+24\eta^{4})}{(1-\eta)^{6}}
\end{equation*}
Using Eqs.~(\ref{cr-point}), (\ref{t_G}) and
(\ref{a1_sw})-(\ref{a3-a4_sw}) one can easily calculate the reduced Ginzburg
temperature. The corresponding results are presented in Table~1 (the first row).  It
should be noted that we get $t_{G}=1.55$ if we pass to the
expression for the the reduced Ginzburg temperature used in  Ref.~\cite{evans} which differs from  Eq.~(\ref{t_G}) by the factor $(\rho_{c}^{*})^{-2}$.
This agrees with the result $t_{G}=1.57$ obtained in  \cite{evans} for the same
system in the RPA.

\subsection{Restricted primitive model}

Now, we consider the RPM setting $\phi^{\text{SR}}\equiv 0$ in Eqs.
~(\ref{a1_rpa})-(\ref{a4_rpa}). Using Eq.~(\ref{wca_coul}),  we get
the following explicit expressions for the coefficients of the
effective Hamiltonian
\begin{eqnarray}
a_{1,0}&=&-\Delta\nu_{N}-\frac{1}{2T^{\text{C}}}+i_{1}, \label{a1_pm} \\
\bar a_{2,0}&=&\frac{(1+2\eta)^{2}}{(1-\eta)^{4}}+i_{2},
\qquad
a_{2,t}=\frac{2\kappa_{c}^{2}}{\pi T_{c}^{C}}\,i_{2,t}, \label{a20_pm} \\
a_{2,2}&=&\frac{\eta(16-11\eta+4\eta^{2})}{20(1-\eta)^{4}}-\frac{1}{6\pi T^{C}}i_{12}, \label{a22_pm}\\
a_{3,0}&=&\frac{(1-7\eta-6\eta^{2})(1+2\eta)}{(1-\eta)^{5}}+i_{3}, \label{a3_pm} \\
a_{4,0}&=&\frac{2(1-6\eta+15\eta^{2}+56\eta^{3}+24\eta^{4})}{(1-\eta)^{6}}+i_{4}.
\label{a4_pm}
\end{eqnarray}
In Eqs.~(\ref{a1_pm})-(\ref{a4_pm}) the following notation are introduced:
\begin{eqnarray*}
i_{n}&=&\frac{(n-1)!(-\kappa^{2})^{n-1}}{\pi T^{C}}
\int_{0}^{\infty} x^{2}\left[\frac{\sin(x)}{x^{3}+\kappa^{2}\sin(x)}\right]^{n}{\rm
 d}x, \nonumber \\
i_{2,t}&=&\int_{0}^{\infty}\frac{x^{5}\sin^{2}(x)}{\left(x^{3}+\kappa_{c}^{2}\sin(x)\right)^{3}}{\rm
 d}x, \nonumber\\
i_{12}&=&\int_{0}^{\infty} x^{6}\left[\kappa^{2}x^{2}(1+\cos^{2}(x)) +2(x^{3}-2\kappa^{2}\sin(x))(2x\cos(x)-3\sin(x))\right. \nonumber \\
&&
\left. +x^{5}\sin(x)\right]/(x^{3}+\kappa^{2}\sin(x))^{4}{\rm
 d}x,
\end{eqnarray*}
where $\kappa=\kappa_{D}\sigma$  with $\kappa_{D}^{2}=4\pi\rho\beta q^{2}z$ being the Debye
 number.
\begin{table}[htbp]
\caption{The reduced  gas-liquid critical parameters, the coefficients of the
effective  Hamiltonian and the  reduced Ginzburg temperature $t_{G}$ for the  hard-core
square-well model (HCSW model) and for the restricted primitive model (RPM) in
the one-loop approximation. The superscripts $\text{SR}$ and
$\text{C}$ refer to the HCSW model and RPM, respectively.}
\vspace{3mm}
\begin{tabular}{ccccccc}
\hline \hline\hspace{5mm}  Model\hspace{6mm} &\hspace{6mm}
$T_{c}^{\text{SR}/\text{C}}$\hspace{6mm} &\hspace{5mm}
$\rho_{c}^{*}$\hspace{8mm}&\hspace{5mm}
$a_{2,t}$\hspace{8mm}&\hspace{5mm}
$a_{2,2}$\hspace{8mm}&\hspace{5mm} $a_{4,0}$\hspace{8mm}&\hspace{5mm}
$t_{G}$\hspace{5mm}
\\
\hline
HCSW  & $1.2667$ & $0.2457$ &  $2.7426$& $0.4536$& $2.7421$  & $0.0937$\\
RPM & $0.08446$ &$0.0088$&  $1.0758$ & $0.2570$ & $0.1752$  & $0.0053$  \\
\hline \hline
\end{tabular}
\end{table}

The reduced temperature $T^{C}$is defined in the standard way  as the ratio between the thermal energy and the interaction strength
of the opposite charged hard spheres at  contact
\begin{equation}
 T^{C}=\frac{k_{\text{B}}T\epsilon\sigma}{zq_{0}^{2}}.
\label{T-C}
\end{equation}
The loci of equations $\bar a_{2,0}=0$ and $a_{3,0}=0$ are shown in
Fig.~1. The two lines  intersect at a maximum of the gas-liquid
spinodal yielding the coordinates of the critical point in the
one-loop approximation (see Table~1).
 \begin{figure}[h]
 \centering
 \includegraphics[height=6cm]{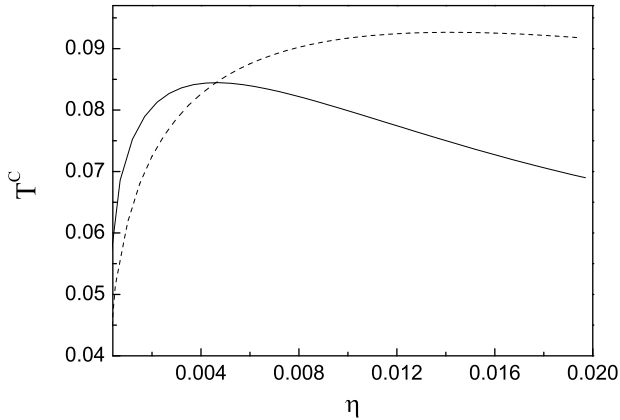}
 \caption{Restricted primitive model: the loci of equations $\bar a_{2,0}=0$
 (solid line) and $a_{3,0}=0$ (dashed line). Temperature $T^{C}$ is given by Eq.~(\ref{T-C}) and
 $\eta=\pi\rho\sigma^{3}/6$ is the packing fraction.} \label{fig1}
 \end{figure}

Substituting the critical temperature and the critical  density into
Eqs~(\ref{a20_pm})-(\ref{a4_pm}) we calculate the coefficients of
the effective Hamiltonian. Then, we get from Eq.~(\ref{t_G}) the  reduced Ginzburg
temperature of the RPM. The results  are presented in Table~1 (the second row). As is
seen, the reduced Ginzburg temperature  found for the RPM is about twenty times
smaller than that  obtained for  the HCSW model. Therefore, contrary
to the previous findings \cite{evans,fisher3,schroer}, our results
suggest that the critical region  for the RPM is much narrower than the
critical region for  a non-ionic model.

\subsection{Long-range interactions versus short-range interactions}

Here,  we consider the full system given by Eq.~(\ref{2_1})  and
study the effect of the interplay of  short-  and long-range
interactions on the Ginzburg temperature. We briefly call the model
as  a RPM-SW model. For the RPM-SW model, we get  explicit
expressions for the coefficients of the effective Hamiltonian
combining Eqs.~(\ref{a1_sw})-(\ref{a3-a4_sw}) and
Eqs.~(\ref{a1_pm})-(\ref{a4_pm}). We also introduce the parameter
\begin{equation*}
\alpha=\frac{T^{SR}}{T^{C}}=\frac{q_{0}^{2}z}{\epsilon\sigma\varepsilon},
\end{equation*}
measuring the strength of the Coulomb interaction with respect to
the solvophobic interaction.
 \begin{figure}[h]
 \centering
 \includegraphics[height=6cm]{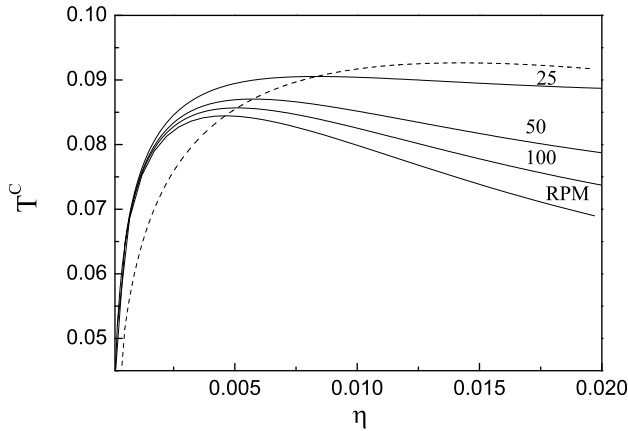}
 \caption{RPM-SW model: the loci of equations $\bar a_{2}=0,0$
 (solid lines) and $a_{3,0}=0$ (dashed line) for $\alpha=25$, $50$, $100$ and
$\infty$ (RPM).  $T^{C}$ is given by Eq.~(\ref{T-C}),
 $\eta=\pi\rho\sigma^{3}/6$ is the packing fraction and $\alpha=q_{0}^{2}z/(\epsilon\sigma\varepsilon)$.} \label{fig2}
 \end{figure}
 \begin{figure}[h]
 \centering
 \includegraphics[height=6cm]{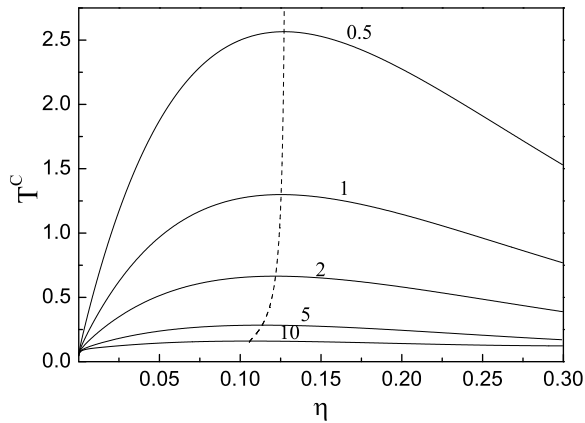}
 \caption{RPM-SW model: the loci of equations $\bar a_{2,0}=0$
 (solid lines) and $a_{3,0}=0$ (dashed line) for $\alpha=0.5,1,2,5$ and $10$. The meaning of the
symbols is the same as in Fig.~2.} \label{fig3}
 \end{figure}

As before,  we first solve the  equations for the critical
parameters. The loci of these equations  are shown in Fig.~2 for
$\alpha=25$, $50$, $100$ and $\infty$ (RPM) and in Fig.~3 for
$\alpha=0.5,1,2,5$ and $10$. As for the RPM, the curve $a_{3,0}=0$
intersects the gas-liquid spinodals at a maximum. The results for
the critical parameters, the  Hamiltonian coefficients and the
reduced Ginzburg temperature are presented in Table~2. As is seen,
the reduced Ginzburg temperature first decreases with an increase of
$\alpha$ and then begins to increase slowly approaching the  RPM
value for  $\alpha\gtrsim 100$. Quite probably, the RPM-SW model
with a certain value of  $\alpha$ lying  between $10$ and $25$
undergoes a tricritical point (see, e.g.,
Ref.~\cite{Ciach_Stell:2002,ciach_stell:2006}). This issue deserves a separate study.
All the coefficients have the same trend with an increase of
$\alpha$ from $0$ to $\infty$, and their trend  coincides with the
trend of $t_{G}$.  Both $T_{c}^{C}$ and $\rho_{c}^{*}$ (or
$\eta_{c}$) decrease when $\alpha$ increases and reach the critical
values of the RPM. The trend of $T_{c}^{SR}$  is opposite to the
trend of $T_{c}^{C}$.

In order to establish a link to our previous results (see Ref.~\cite{patsahan-caillol-mryglod:07}) we pass
from $\alpha$ to the ionicity ${\cal I}$ by means of the relation
\begin{equation}
 \alpha=T^{SR}{\cal I}, \qquad \text{where}  \qquad
{\cal I}=\displaystyle\frac{1}{T^{C}}.
\label{Ion}
\end{equation}
Fig.~4 presents  the dependence of the reduced Ginzburg temperature
on the ionicity. The results of the present work are shown by solid
circles while the line represents the results  from
Ref.~\cite{patsahan-caillol-mryglod:07}. As is seen,   the  both
groups of results agree well for  ${\cal I}\leq 6.25$ corresponding
to $\alpha \leq 10$. The deviations between the results that appear
for  large values of ${\cal I}$  are due to the different
approximations  used in  calculating  the Hamiltonian coefficients
in the present and previous works. Therefore, the results of this
paper generally agree with the  previous results obtained within the
framework of a different perturbative treatment and, in turn,
confirm a key role of the Coulomb interactions in the reduction of
the crossover region  compared to the solvophobic interactions.

%
\begin{table}[htbp]
\caption{The reduced gas-liquid critical parameters, the
coefficients of the effective Hamiltonian and the  reduced Ginzburg
temperature $t_{G}$ for the  RPM-SW model (see the text) in the
one-loop approximation. } \vspace{3mm}
\begin{tabular}{cccccccc}
\hline \hline\hspace{5mm}  $\alpha$\hspace{5mm} &\hspace{5mm}
$T_{c}^{\text{C}}$\hspace{8mm} &\hspace{8mm}
$T_{c}^{\text{SR}}$\hspace{5mm}&\hspace{5mm}
$\rho_{c}^{*}$\hspace{7mm}&\hspace{5mm}
$a_{2,t}$\hspace{6mm}&\hspace{5mm}
$a_{2,2}$\hspace{8mm}&\hspace{5mm} $a_{4,0}$\hspace{8mm}&\hspace{5mm}
$t_{G}$\hspace{5mm}
\\
\hline $0.5$& $2.56537$ & $1.28269$ &$0.2429$& $2.7132$ & $0.4506$&
$2.6606$& $0.0903$
\\
$1$& $1.29904$ & $1.29904$ &$0.2393$& $2.6694$ & $0.4427$& $2.5441$&
$0.0885$
\\
$2$& $0.66471$&$1.32943$ &$0.2328$&$2.5908$&$0.4288$ &$2.3275$&$0.0840$ \\
$5$ &$0.28387$ &$1.41935$ &$0.2170$&  $2.4320$ & $0.3986$ & $1.8133$ & $0.0676$  \\
$10$ &$0.16012$ & $1.6012$&$0.2026$&  $2.4104$ & $0.3740$ & $1.2586$ & $0.0390$  \\%
$25$ & $0.09054$& $2.2636$ &$0.0159$&  $1.0406$ & $0.2175$ & $0.0906$  & $0.0024$  \\
$50$&$0.08705$ & $4.3524$&$0.0109$  & $1.0597$& $0.2396$& $0.1472$  & $0.0047$  \\
$100$ & $0.08568$& $8.568$ & $0.0097$ &  $1.0676$& $0.2481$& $0.1602$  & $0.0050$\\
\hline \hline
\end{tabular}
\end{table}

\begin{figure}[h]
\centering
 \includegraphics[height=6cm]{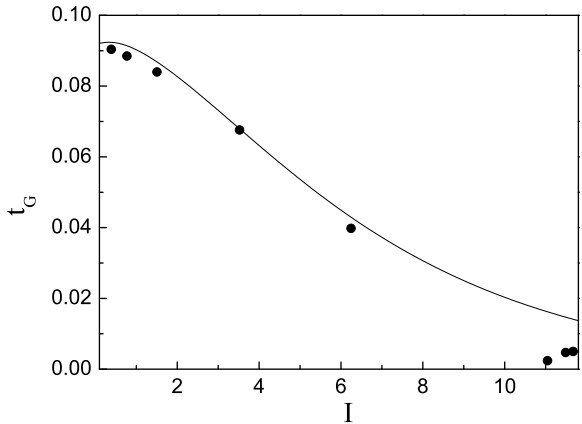}
 \caption{The reduced  Ginzburg temperature as a function of the ionicity (see the text).
 The symbols indicate the  results
of the present work and the    line represents  the results from Ref.~\cite{patsahan-caillol-mryglod:07}.
${\cal I}$ is given by Eq.~(\ref{Ion}).} \label{fig4}
 \end{figure}

\section {Conclusions}
In this paper, we have revisited the issue of the criticality of the
Coulomb dominated systems.  The model considered,
besides Coulomb interactions, includes  short-range attractive
interactions. Using  the CVs-based theory we have
derived the $\phi^{4}$-model  LGW Hamiltonian  in terms of the CVs describing the
fluctuation modes of the total number density.
The resulting form of the effective
Hamiltonian  confirms the fact that the criticality of the Coulombic models
belongs to the universal class of a three dimensional Ising
model.
The important feature of the developed approach  is that it enables  us to obtain  all the
coefficients, including the square-gradient term,  within the framework  of
the same approximation.

 The LGW Hamiltonian obtained   has been used for the calculation of the gas-liquid
critical parameters and the Ginzburg temperature for a number of
models, in particular,  the pure solvophobic model, the pure
Coulombic model as well as for the models including both the Coulomb
interactions and the  solvophobic interactions. In the present
paper, all quantities, the critical parameters, the relevant
coefficients of the effective Hamiltonian, and, accordingly, the
Ginzburg temperature, for all the models considered  have been
calculated in the one-loop approximation. This approximation leads
to the mean-field values for the  reduced critical parameters. It
is worth noting that the charge-asymmetry does not manifest itself
at this level of approximation.

Having excluded the Coulomb interaction, we first calculate the
reduced Ginzburg temperature for a  simple fluid model, the
one-component system of hard spheres  interacting via the
square-well potential of the range $\lambda=1.5$. This well-known
model is used to calibrate the Ginzburg temperature for the models
having Coulomb interactions. Then, neglecting the short-range
attraction we study the criticality in the RPM. Our calculations
have shown that the reduced Ginzburg temperature of the Coulombic systems
is much smaller (although it is not extremely small) than for
simple fluids.
We have obtained the fluctuation dominated region for  the RPM
to be much narrower  (reduced by a factor of $\sim 20$) than
for the HCSW fluid.

We have also studied  the Ginzburg temperature depending on the
interplay between the Coulombic and solvophobic interactions. Having
introduced the ratio $\alpha$ that determines the  strength of the
Coulomb interaction with respect to the  solvophobic interaction (SW
potential), we  calculate both the reduced critical temperature and
critical density, the Hamiltonian coefficients and the reduced
Ginzburg temperature for a set of the $\alpha$ values ranging from
$0.5$ to $100$. For the model under consideration,  the Ginzburg
temperature  shows a non-monotonous behaviour with the variation of
$\alpha$. In particular, $t_{G}$ first decreases approaching its
minimum in the region $10<\alpha<25$ and then starts to increase
approaching the RPM value for $\alpha>100$.  The
present results generally agree  with our previous  results obtained
within the framework of the different perturbative treatment.
Moreover, they are in  good quantitative agreement  for  $\alpha\leq
10$.

In conclusion,   within the framework of the unified approach we
have  derived the Ising-like effective Hamiltonian which enables  us
to make  reliable estimates of the reduced Ginzburg temperature for
the RPM as well as for the model with the competing solvophobic and
Coulomb interactions.  For the model including both the short-range
attractive interactions and the long-range Coulomb interactions, we
have shown that the reduced Ginzburg temperature approaches the value 
obtained for the RPM  when the strength of the
Coulomb interactions becomes large enough.  This suggests the key
role of the Coulomb interactions in the crossover behaviour observed
experimentally  in ionic fluids.   We believe that our results
provide new insights into the nature of the non-classical region in
the Coulomb-dominated systems.

\section{Appendix}

\subsection{Expressions for  coefficients $a_{n}^{(i_{n})}$}
Taking into account the recurrence formulae for
${\mathfrak{M}}_{n}^{(i_{n})}$ (see~(46) in Ref.~\cite{Pat-Mryg-CM})
we get the following expressions for coefficients $a_{n}^{(i_{n})}$
for $n\leq 4$
\begin{equation*}
a_{2}^{(0)}=\beta\rho\widetilde{\phi}^
{SR}(k)+\frac{1}{S_{2,\text{HS}}(k)}, \qquad
a_{2}^{(2)}=\beta\rho\widetilde{\phi}^ {C}(k)+\frac{1}{q_{0}^{2}z},
\end{equation*}
\begin{equation*}
a_{3}^{(0)}=-\frac{S_{3,\text{HS}}}{S_{2,\text{HS}}^{3}}, \qquad
a_{3}^{(2)}=-\frac{1}{q_{0}^{2}z}, \qquad
a_{3}^{(3)}=-\frac{(1-z)}{q_{0}^{3}z^{2}},
\end{equation*}
\begin{eqnarray*}
a_{4}^{(0)}&=&-\frac{1}{S_{2,\text{HS}}^{4}}\left(S_{4,\text{HS}}-\frac{3S_{3,\text{HS}}^{2}}{S_{2,\text{HS}}}\right),
\qquad a_{4}^{(2)}=\frac{2}{q_{0}^{2}z},
\nonumber \\
a_{4}^{(3)}&=&\frac{2(1-z)}{q_{0}^{3}z^{2}}, \qquad
a_{4}^{(4)}=-\frac{2(1-z+z^{2})}{q_{0}^{4}z^{3}},
\end{eqnarray*}
where $S_{n,\text{HS}}=\widetilde G_{n,\text{HS}}/\langle
N\rangle_{\text{HS}}$, $\widetilde G_{n,\text{HS}}$ is the Fourier
transform of the $n$-particle connected correlation function of a
one-component hard-sphere system.

In order to get  the corrections $\Delta a_{3}$ and $\Delta a_{4}$
(Eqs.~(\ref{a3})-(\ref{a4}))  we  also need coefficients
$a_{5}^{(2)}$ and $a_{6}^{(2)}$ presented by
\begin{equation*}
 a_{5}^{(2)}=-\frac{3!}{q_{0}^{2}z},
\qquad a_{6}^{(2)}=\frac{4!}{q_{0}^{2}z}.
\end{equation*}

\end{document}